\documentclass[letterpaper, 10 pt, conference]{ieeeconf}  

\IEEEoverridecommandlockouts                              

\usepackage{graphics} 
\usepackage{amsmath} 
\usepackage{amssymb}  
\usepackage{algorithmic}
\usepackage{algorithm}
\usepackage{mathtools}
\usepackage{subfigure}

\title{\LARGE \bf
Towards Integrated Traffic Control \\ with Operating Decentralized Autonomous Organization
}

\author{Shengyue Yao$^{1}$, Jingru Yu$^{1}$, Yi Yu$^{1}$, Jia Xu$^{1}$, Xingyuan Dai$^{2}$, \\Honghai Li$^{3}$,
Fei-Yue Wang$^{4}$, \textit{Fellow, IEEE}, Yilun Lin$^{1,*}$, \textit{Member, IEEE}  
\thanks{*This work was supported by the Shanghai Artificial Intelligence Laboratory.}
\thanks{*Corresponding author: Yilun Lin (linyilun@pjlab.org.cn)} 
\thanks{$^{1}$Shengyue Yao (yaoshengyue@pjlab.org.cn), Jingru Yu (yujingru@pjlab.org.cn), Yi Yu (yuyi@pjlab.org.cn), Jia Xu (xujia@pjlab.org.cn) and Yilun Lin (linyilun@pjlab.org.cn) are with Urban Computing, Shanghai AI Lab, Shanghai, China.}
\thanks{$^{2}$Xingyuan Dai (xingyuan.dai@ia.ac.cn) is with the The State Key Laboratory for Management and Control of Complex Systems, Institute of Automation, Chinese Academy of Sciences, Beijing, 100190, China.}%
\thanks{$^{3}$Honghai Li (honghai\_1@126.com) is with the Research and Development Center of Transport Industry of Autonomous Driving Technology, RIOH High and Technology Group.}%
\thanks{$^{3}$Fei-Yue Wang (feiyue.wang@ia.ac.cn) is with the Institute of Automation, Chinese Academy of Sciences, Beijing, China, and the Macau Institute of Systems Engineering, Macau University of Science and Technology, Macau, China.}
}

\begin{document}

\maketitle
\thispagestyle{empty}
\pagestyle{empty}

\begin{abstract}
With a growing complexity of the intelligent traffic system (ITS), an integrated control of ITS that is capable of considering plentiful heterogeneous intelligent agents is desired. However, existing control methods based on the centralized or the decentralized scheme have not presented their competencies in considering the optimality and the scalability simultaneously. To address this issue, we propose an integrated control method based on the framework of Decentralized Autonomous Organization (DAO). The proposed method achieves a global consensus on energy consumption efficiency (ECE), meanwhile to optimize the local objectives of all involved intelligent agents, through a consensus and incentive mechanism. Furthermore, an operation algorithm is proposed regarding the issue of structural rigidity in DAO. Specifically, the proposed operation approach identifies critical agents to execute the smart contract in DAO, which ultimately extends the capability of DAO-based control. In addition, a numerical experiment is designed to examine the performance of the proposed method. The experiment results indicate that the controlled agents can achieve a consensus faster on the global objective with improved local objectives by the proposed method, compare to existing decentralized control methods. In general, the proposed method shows a great potential in developing an integrated control system in the ITS. 
\end{abstract}

\section{INTRODUCTION}

The development of the intelligent traffic system (ITS) has long been focused in academia and industry over the past decades, which has evolved into a complex system with plentiful intelligent agents, including connected and automated vehicles (CAV) and multiple intelligent traffic control measures (e.g., traffic signal, variable speed limits, perimeter gating, etc.). It is foreseeable that the ITS will become increasingly complex with the surge of intelligent agents, thus the control of ITS becomes increasingly essential. Specifically, the control of ITS should consider all involved intelligent agents which are heterogeneous in different scales, such as in different responding frequency; as well as with diverse objectives, such as mitigating congestion, guaranteeing safety, reducing emission and energy consumption. 

Existing optimal control methods of individual agents in ITS are theoretically mature and have evolved towards more expeditious, agile, and adaptive with the aid of AI-based approaches \cite{han2022new,li2016traffic,belletti2017expert}. However, given the complex nature of ITS, these methods are incapable of performing as expected in practice, while a wide deployment of intelligent agents has frequently trapped the entire ITS into the Braess paradox \cite{diakaki2015overview,wolpert2002collective}. Therefore, investigating an integrated control method, which is capable of coordinating the behaviours of heterogeneous agents in ITS renders a non-trivial task to be tackled.     

Recent research focuses on enhancing scalability and optimality in integrated traffic control, for which a centralized or decentralized control scheme is adopted, respectively. The centralized scheme assumes perfect knowledge of all agents' states and optimizes global objectives, while the decentralized scheme optimizes local objectives and achieves Pareto optimality\cite{baskar2006decentralized}. However, the centralized control scheme has its limit in coordinating heterogeneous agents, whereas the decentralized scheme is inefficient in synchronizing agents' behaviours. Therefore, neither the centralized nor the decentralized scheme presents its competency in the integrated control of plentiful heterogeneous intelligent agents.  

Fortunately, the emergence of Blockchain technology and Decentralized Autonomous Organization (DAO) provide opportunities for developing integrated traffic control with scalability and optimality considerations\cite{miaoDAOHANOIDeSci2023,mollahBlockchainInternetVehicles2021}. DAO controls a complex system of heterogeneous agents using smart contracts, ensuring control optimality through a proposal-voting-action-incentive mechanism. In addition, the autonomous and independent execution of smart contracts ensures scalability and secures data transmission. While the potential of DAO-based control in ITS has been discussed, few studies have evaluated its performance in practice. Specifically, the feasible design of smart contracts and endogenous governing defects of DAO, especially the structural rigidity\cite{chohan2017decentralized}, pose challenges in its application in practice, particularly in complex systems with numerous intelligent agents.

With the consideration of the issues above, a consensus and incentive mechanism for DAO-based ITS control and an operation on DAO are proposed in this research. Specifically, the proposed mechanism focuses on optimizing the energy consumption efficiency (ECE) in ITS as mentioned by Wang \cite{wangDAOMetaControlMetaSystems2022}. The ECE represents the rate of improved local objectives and changed control effort magnitude, which should be maximized and balanced through all involved agents. Meanwhile, the operation on DAO unveils critical agents in ITS to execute smart contracts in DAO, in order to avoid frequent code altering in deployed smart contracts or deploying new smart contracts. In addition, the operation on DAO is inspired by recent studies in the dense reinforcement learning and the message transaction optimization for Blockchain-enabled Internet-of-Vehicles \cite{fengDenseReinforcementLearning2023,sharma2018energy}.

The rest of this paper is organized as follows. Section \ref{related work} investigates the related work regarding existing integrated control schemes and DAO-based control mechanism. Section \ref{methodology} elaborates the methodology including the problem formulation, the solving framework, the detailed algorithms of the operation on DAO and the consensus and incentive mechanism. Section \ref{case study} discusses the application of the proposed integrated control method in ITS, as well as  presents examination results of the proposed method in a numerical experiment. Finally, this research is concluded in section \ref{conclusion}

\section{RELATED WORK}
\label{related work}
\subsection{Integrated traffic control in ITS}
In this section, existing integrated control schemes in ITS (centralized and decentralized schemes) are reviewed. Specifically, the application scenario and the defects of centralized and decentralized control scheme are elaborated, respectively.
\subsubsection{Centralized control method}
The centralized scheme operates under the assumption of a perfect knowledge of all controlled agents' states. This knowledge is then leveraged to regulate the agents' behaviors, optimizing a global objective that encapsulates the objectives of all agents. 

In ITS, the centralized control scheme is commonly employed in managing multiple homogeneous agents, such as CAV platoon control \cite{yao2019managing,milanes2013cooperative,shi2021connected}, traffic signal coordination \cite{liu2020receding}, and highway control measures coordination \cite{roncoli2015traffic,fares2022marl}.
However, the centralized scheme exhibits limitations in scalability and is incompetent in coordinating heterogeneous intelligent agents. Additionally, the assumption of perfect knowledge raises various concerns about communication reliability and security \cite{shet2021cooperative,raja2022blockchain}.
\subsubsection{Decentralized control method}
The decentralized scheme operates by regulating intelligent agents based on local information as well as information received from their connected agents. This approach optimizes their behaviors towards their local objectives while achieving a consensus to achieve a Pareto optimality.

In ITS, the decentralized scheme is frequently employed in coordinating behaviors of heterogeneous agents, such as coordinating multiple urban traffic control measures with diverse objectives (e.g., mitigating traffic congestion and reducing evacuation time after an event \cite{wang2010parallel}) or in different scales \cite{gerostathopoulos2019trapped,li2020perimeter}. However, the defects in  achieving optimization and synchronization impedes the decentralized scheme from being adopted to regulate a large number of intelligent agents \cite{baskar2006decentralized, fares2022marl}.

\begin{figure}[thpb]
\centering
\includegraphics[width=3.0in]{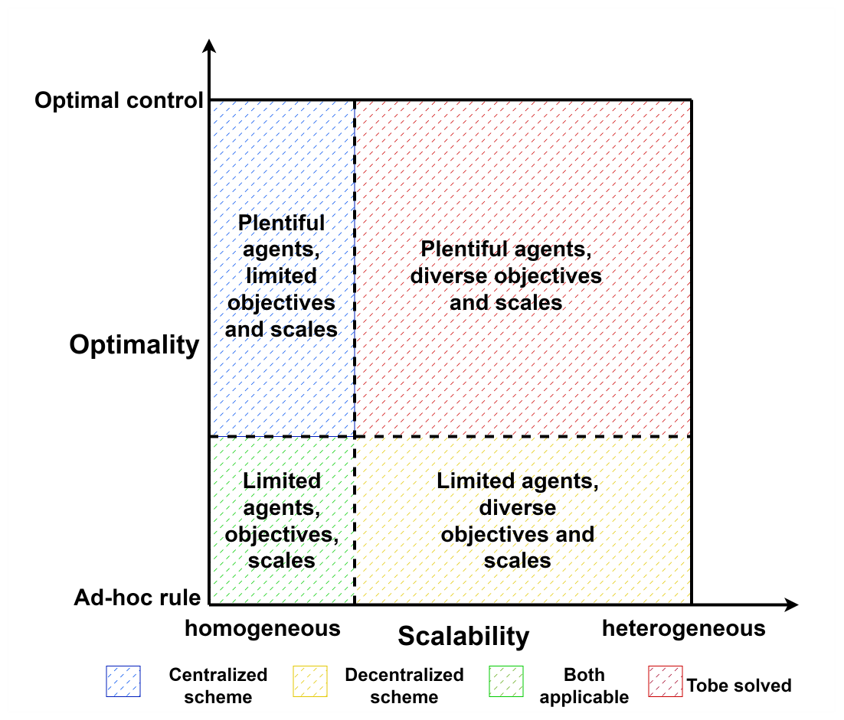}
\caption{Existing integrated control scheme capability and unsolved issue}
\label{capability}
\end{figure}

In summary, neither the centralized nor the decentralized scheme has demonstrated its competence in the integrated control of numerous heterogeneous intelligent agents. An integrated ITS control which can achieve both the optimality and the scalability is desired, as it is illustrated in Fig. \ref{capability}.

\begin{figure}[thpb]
\centering
\includegraphics[width=3.0in]{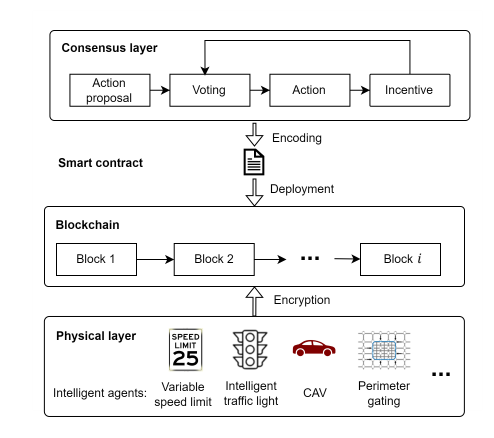}
\caption{DAO-based control mechanism}
\label{dao-based control frame}
\end{figure}

\subsection{DAO and related control methods}

The emergence of Blockchain technology and decentralized autonomous organization (DAO) provides a promising avenue for developing an integrated traffic control system that considers both the scalability and the optimality. It is worth noting that the deployment of smart contracts on the Blockchain can further guarantee secure and private-preserving data transmission, although this aspect is not discussed in this study \cite{hassanDecentralizedAutonomousOrganization2021,wang2019decentralized}.

The decentralized and autonomous nature of the DAO has spurred interests in applying it to the integrated control of ITS. \cite{miaoDAOHANOIDeSci2023,mollahBlockchainInternetVehicles2021}. In addition, the framework of Dao-based control in ITS has been extensively discussed, which can be used in integrated control of urban traffic system \cite{wang2019decentralized}, crowd-sensing systems \cite{zhuCrowdsensingIntelligenceDecentralized2022}, vehicle parking recommendation systems\cite{renIntelligentDesignImplementation2019}, etc. In general, agents in physical world are enecrypted into Blocks in Blockchians, which can automatically and independently execute smart contracts that are encoded and deployed in advance. Moreover, the smart contracts stimulates agents to achieve a consensus through a proposal-voting-action-incentive mechanism \cite{bellavitisRiseDecentralizedAutonomous2023}. The DAO-based control framework can be summarized in Fig. \ref{dao-based control frame}. 

However, the detailed design of the smart contract in ITS control has received little attention, particularly the formulation of the consensus and incentive mechanism. Another critical issue is the structural rigidity. Specifically, altering the code in the smart contract once the system is operating can be extremely difficult, while deploying new smart contracts can be expensive \cite{chohan2017decentralized,dupontExperimentsAlgorithmicGovernance}. The issue of structural rigidity poses a significant challenge when applying the DAO in a complex system with a growing number of intelligent agents, such as in ITS.

Despite extensive research on integrated traffic control, the research gap still lies in the lack of a suitable control design that can achieve both scalability and optimality while regulating  numerous heterogeneous intelligent agents. Particularly, the detailed DAO-based control design addressing the formulation of the consensus and incentive mechanism and the structural rigidity issue is desired. 

%
%

\section{METHODOLOGY} 
\label{methodology}


\subsection{Problem definition}
\label{definition}
The problem of an integrated ITS control, considering plentiful heterogeneous intelligent agents, is defined in this section. 

The problem is defined under the following assumptions: (1) The intelligent agents in ITS are connected through Peer-to-Peer (P2P) communication and the communication delay is negligible (in milliseconds) compare to the control response time (in seconds). (2) The connection intensities, which reflect the exchange rates of control effort between agents, are known a-priori through learning the history data.  

Based on the assumptions above, we focus on developing a DAO-based integrated control, which aims to increase and balance the ECE of all intelligent agents in ITS, upon the optimization of their local objectives.

For a road network contains a set of $N$ intelligent agents $V_N = \{v_1,...v_n\}$ with different response time $\tau_i$, its communication topology can be represented by a graph $G$, that $G=\{V_N, E, A_{N \times N}\}$. $E \subseteq V_N \times V_N$ denotes the set of edges $e_{ij}$, which indicates the communication exists between agents $i$ and $j$, $i \subseteq N$, $j \subseteq N$. $A_{N \times N}$ is the adjacent matrix, that $a_{ij} \subseteq \mathbb{R}, \forall e_{ij} \subseteq E$ and $a_{ij}=0$ otherwise. In addition, $a_{ii}=0$. The value of $a_{ij}$ indicates the linking intensity, that $a_{ij} = \frac{u_j}{u_i}$. The dynamic of agent $v_i$ can be represented by an ODE in Eq. (\ref{ODE}).
\begin{subequations}
\label{ODE}    
\begin{equation}
\Dot{x_i}=f(x_i,\textbf{u}_i)   
\end{equation}
\begin{equation}
\textbf{u}_i = u_i + \sum_{j=1}^{N}a_{ji}u_j
\end{equation}
\end{subequations}
where $x_i$ is the observed state of agent $v_i$, and the control input $u_i$ is optimized towards its local objective $J_i=\min g(x_i, u_i)$. For the integrated control of the defined road network system, all intelligent agents in the road network are controlled towards their local objectives. Meanwhile, a global objective of maximizing the cumulative ECE, meanwhile achieving a consensus on the ECE, is desired, as represented by Eq. (\ref{global objective}).

\begin{subequations}
\label{global objective}
\begin{equation}
\label{cummulative obj}
J_G^{cumulative} = \max \int_{t=0}^\infty \sum_{i}^{N}r_{i}
\end{equation}
\begin{equation}
\label{balance obj}
J_G^{balance} = \lim_{t \to \infty}|  r_i - r_j | = 0, \forall i\subseteq N, \forall j \subseteq N
\end{equation}
\text{s.t.}
\begin{equation}
r_{i} = \frac{\Dot{g_i}}{|\Dot{u_i}|}
\end{equation}
\end{subequations}
where $|\cdot|$ denotes the absolute value. Note that the ECE of an agent $r_i$ represents the rate of its changed local objectives and its changed control effort magnitude. Hence a higher value of ECE indicates that the agent is able to achieve a better performance with less consumed energy in changing its control effort. Therefore, the ECE should be maximized and balanced among all agents in the controlled system with an integrated control, as indicated by Eq. (\ref{global objective})  

As it is mentioned in section \ref{related work}, solving the defined problem in a centralized scheme is intricate due to the heterogeneity of agents, as well as the difficulties in summarizing Eq. (\ref{cummulative obj}) and Eq. (\ref{balance obj}). In addition, the defined problem cannot be effectively solved in a decentralized scheme due to the synchronization issue. Even with the DAO-based control, the issue of structural rigidity with the growing size of set $A$ and $E$ posses a critical challenge. Therefore, instead of analytically solving the defined problems, a solving method based on reiteratively executing a DAO-based control protocol and operating the DAO is proposed.
\subsection{Solving framework}
\begin{figure}[!t]
\centering
\includegraphics[width=3.0in]{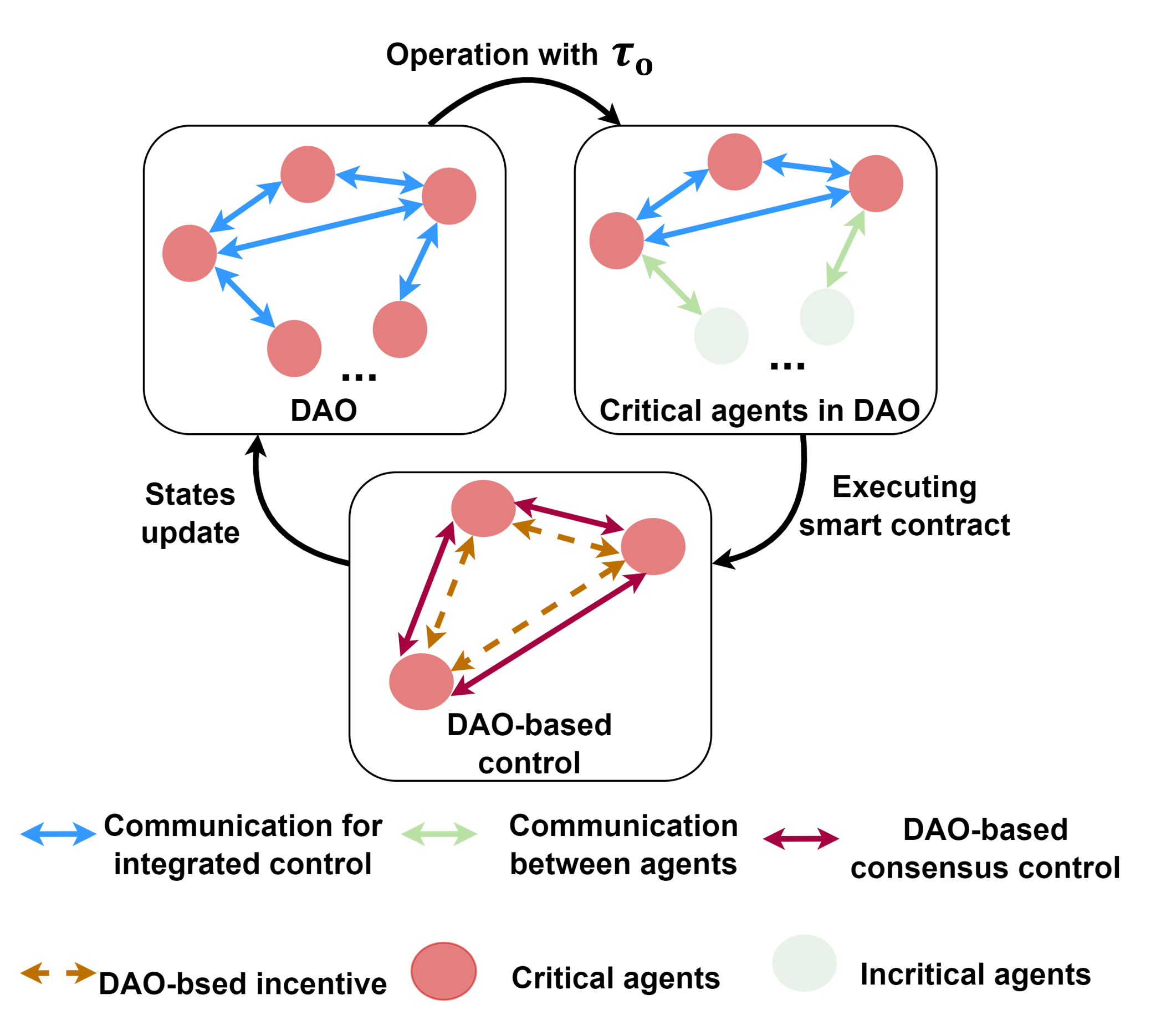}
\caption{Problem solving framework}
\label{dao-based proposed framework}
\end{figure}

Addressing the issues of heterogeneity, synchronization and DAO's structural rigidity, the problem defined in section \ref{definition} can be solved by a framework as in Fig. \ref{dao-based proposed framework}. In general, the operation on DAO extracts a sub-graph $G^s=\{V^s_{N^s}, E^s_{N^s}, A^s_{N^s \times N^s}\}$ from $G$ with a frequency of $\tau_{o}$ by identifying critical agents, which will be elaborated in section \ref{operation on DAO}. Afterwards, a DAO-based control protocol encoded in the smart contract will be executed in $G^s$. Specifically, the DAO-based control considers consensus among all involved agents (Eq. \ref{balance obj}) and improving local objectives. Meanwhile, the global objective in Eq. (\ref{cummulative obj}) can be improved by distributing incentives, which will be elaborated in section \ref{consensus and incentive mechanism}.

\subsection{Operation on DAO}
\label{operation on DAO}


According to Fig. \ref{dao-based proposed framework}, the operation on DAO aims to identify critical agents by extracting $G^s$ from $G$, in order to avoid deploying new smart contracts to accommodate the growing amount of agents in DAO. Given the deployed smart contract capability $\Phi$ (i.e., the amount of intelligent agents in ITS that the deployed smart contract is capable of effectively executing an integrated control), and the tolerance level $\Psi$ (i.e. the magnitude of global optimization objective that can be compromised during the operation), the objective of the operation on DAO can be formulated in Eq. (\ref{densed DAO objective}).


\begin{subequations}
\label{densed DAO objective}
\begin{equation}
    J^s = \min (|E^s_{N^s}|)
\end{equation}
\text{s.t.}
\begin{equation}
    W_i(G^s) \leq \Psi, \forall i \subseteq N^s
\end{equation}
\begin{equation}
    W_i(G^s) = \sum_{j \subseteq N} a_{ij} \cdot |r_i-r_j| - \sum_{j \subseteq N^s} a_{ij} \cdot |r_i-r_j| 
\end{equation}
\begin{equation}
    N^s \leq \Phi 
\end{equation}
\end{subequations}

A heuristic algorithm is proposed in Algorithm \ref{DAO operation algorithm} to find an approximate solution for achieving Eq.(\ref{densed DAO objective}). Note that if a feasible $G^s$ is not found by Algorithm \ref{DAO operation algorithm}, the proposed operation on the existing DAO may reach its capacity limit and new smart contracts are necessary to be deployed.
\begin{algorithm}[h]
\caption{DAO operation algorithm}\label{DAO operation algorithm}
\begin{algorithmic}[1]
\REQUIRE $G$, $\Phi$,$\Psi$
\ENSURE $G^s$
\STATE Initialization $G^s = G$, $N^s = N$, $\textbf{N}^s=\{1,2,...N\}$
\WHILE{Maximum iteration is not reached}
    \WHILE{$N^s > \Phi$}
        \STATE Randomly select $i \subseteq \textbf{N}^s$
        \IF{$\forall j, e_{ij} \notin E^s$}
            \STATE $\textbf{N}^s=\textbf{N}^s/\{i\}$
            \STATE $N^s = N^s-1$
        \ENDIF
        \STATE $w_i = \min(a_{ij} \cdot |r_i - r_j|)$, $\forall j$ that $e_{ij} \subseteq E^s$
        \STATE $\Tilde{G^s}=\{V^s, E^s/\{e_{ij}\}, \{A^s:a_{ij}=0\}\}$
        \IF{$W_i(\Tilde{G}) \leq \Psi$}
            \STATE $G^s=\Tilde{G^s}$
        \ELSE
            \STATE \textbf{break}
        \ENDIF
    \ENDWHILE
    \STATE \textbf{return} $G^s$
\ENDWHILE  
\STATE \textbf{return} did not find $G^s$
\end{algorithmic}  
\end{algorithm}

\subsection{Consensus and incentive mechanism}
\label{consensus and incentive mechanism}
Having the critical agents identified with $G^s$, an integrated control of $v_i \subseteq V_i^s$ is achieved through a consensus and incentive mechanism. Initially, a control protocol is designed in Eq. (\ref{control protocol}), which is inspired by the protocol formulated by Xie et al. \cite{xie2017global}. 

\begin{subequations}
    \label{control protocol}
    \begin{equation}
    u_i = \sigma_{\Delta}(-d_i-\alpha_i \bigtriangledown_{u_i}g(x_i,u_i)-\beta_i \sum_{j=1} ^{N^s} a_{ij}(r_i-r_j))        
    \end{equation}
    \text{s.t.}
    \begin{equation}
    \Dot{d_i} = \alpha_i \beta_i \sum_{j=1} ^{N^s} a_{ij}(r_i-r_j), d_i(0) = 0        
    \end{equation}
    \begin{equation}
    \sigma_{\Delta}(u) = \text{sgn}(u) \min \{|u|, \Delta\} 
    \end{equation}
\end{subequations}
where $N^s$ is the amount of agents in $V^s$, $\bigtriangledown_{u_i}g(x_i,u_i)$ is the local objective optimization term, $\sum_{j=1} ^{N^s} a_{ij}(\frac{r_i}{\tau_i}-\frac{r_i}{\tau_i})$ is the consensus term, and $-d_i$ is the stabilizing term that maintains the consensus at the optimal point of local objectives. $\sigma_{\Delta}$ is the saturation function bounding the value of control input with a level $\Delta$, we set $\Delta = \min\{|u_i^{min}|, |u_i^{max}|\}$ for simplicity. In addition, $\alpha_i, \beta_i \subseteq \mathbb{R}^{+}$ are weighting coefficients with respect to local and global objectives, respectively. 

According to the framework of DAO-based control in Fig. \ref{dao-based control frame}, the value of $\alpha_i$ and $\beta_i$ represent the voting power of agent $i$, which can be proportional to the mining workload \cite{al2019privacy}, the reputation of following the traffic rules\cite{wang2019bsis}, or the discrepancy between its current state and the consensus goal\cite{zhuBlockchainbasedConsensusStudy2021}. In this study, the Proof-of-Stack (PoS) scheme by Zhu et al. \cite{zhuBlockchainbasedConsensusStudy2021} is adopted and the value of $\alpha_i$ and $\beta_i$ are defined by Eq. (\ref{voting form}).

\begin{subequations}
    \label{voting form}
    \begin{equation}
        \alpha_i = \gamma_i \exp({k_1 \Tilde{r_{i}}})
    \end{equation}
    \begin{equation}
        \beta_i = \gamma_i \exp{-(k_2 \Tilde{r_{i}})}
    \end{equation}
    \text{s.t.}
    \begin{equation}
        \Tilde{r_{i}} = \sum_{j=1}^{N^s}a_{ij} (r_i - \frac{r_i+r_j}{2})
    \end{equation}
\end{subequations}
$k_1, k_2 \subseteq \mathbb{R}^{+}$ are positive constants. $\gamma_i \subseteq \mathbb{R}^{+}$ is the weighting coefficient. Following the process of DAO-based control, $\gamma_i$ updates according to the incentives received by agent $i$ after $u_i$ is executed, that:
\begin{subequations}
\begin{equation}
    \label{rewards}
    \gamma_i^{update} = \gamma_i^{old} (\frac{2*\arctan(h_i)}{\pi}+1)
\end{equation}
\text{s.t.}
\begin{equation}
     h_i = k_3 \cdot (\Tilde{r_i}-\Tilde{r_i}^{old}) + k_4 \cdot \sum_{j=1}^{N^s}a_{ij} (g_j-g_j^{old})
\end{equation}
\end{subequations}

$k_3$,$k_4 \subseteq \mathbb{R}^{+}$ are positive constants. The term $k_3 \cdot (\Tilde{r_i}-\Tilde{r_i}^{old})$ presents the reward from achieving a consensus, whereas the term $k_4 \cdot \sum_{j=1}^{N^s}a_{ij} (g_j-g_j^{old})$ presents the 'support' or 'objection' from the connected agents of $v_i$ in $V^s$. 



\begin{figure}[thpb]
\centering
\includegraphics[width=2.5in]{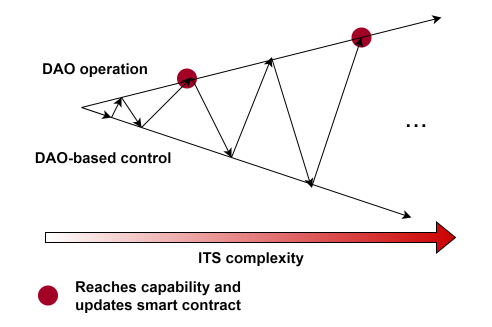}
\caption{Integrated control development process}
\label{DAO-based control process}
\end{figure}

With the proposed solving method, the development of an integrated control in ITS can follow the process of repeatedly deploying smart contract and operating DAO with the growth of ITS system complexity, as presented in Fig.\ref{DAO-based control process}. Following this process, the capability of DAO-based ITS control in achieving optimality and scalability can be extensively enhanced.  

\section{CASE STUDY AND RESULTS DISCUSSION}
\label{case study}
\subsection{Application scenario in ITS}
\begin{figure}[t]
    \centering
    \subfigure[A typical scenario of integrated control in ITS by the proposed method]
    {
        \includegraphics[width=0.8\linewidth]{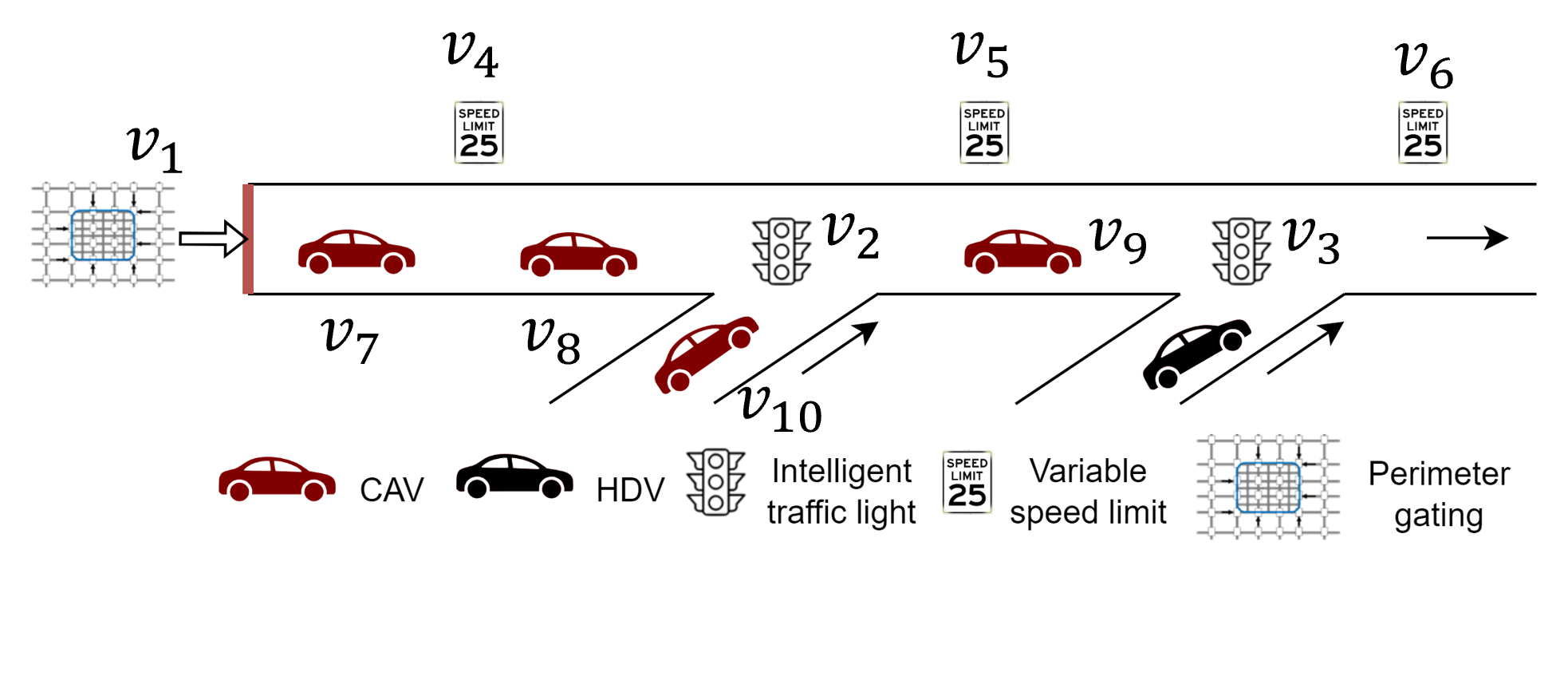}
        \label{scenario}
    } 
        \subfigure[Integrated control framework in the described scenario]
    {
        \includegraphics[width=0.8\linewidth]{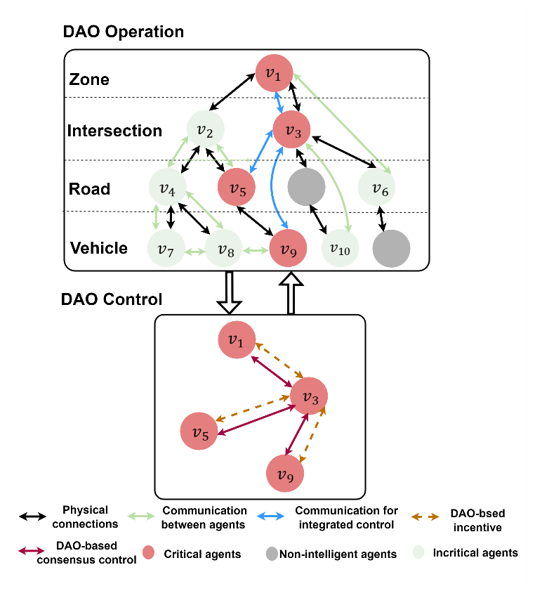}
        \label{scenario-dao}
    }
    \caption{Application scenario in ITS}
    \label{ITS scenario}
\end{figure}

Fig. \ref{scenario} illustrated a typical scenario that can apply the proposed integrated control method. Specifically, a single-lane arterial road with two three-arm intersections is depicted. There exists multiple intelligent agents in the depicted scenario, including CAVs, intelligent traffic lights, variable speed limits (VSL), perimeter gating, etc, which exhibits significant heterogeneity in terms of their scales and objectives. For instance, the acceleration of a CAV can be decided in milliseconds, with the objectives of reducing travel delay and acceleration oscillation; whereas the control of gated flow by the perimeter gating control aims to decrease the total travel delay of vehicles in the controlled network, with a response time in minutes.

With the proposed method, intelligent agents can be controlled via reiteratively operating DAO and executing a DAO-based control through a consensus and incentive mechanism, as presented in Fig. \ref{scenario-dao}. It is worthwhile to note that except for addressing the issue of heterogeneity and synchronization, the proposed method has practical significance. For instance, if CAV $v_9$ encounters an emergency, the agents $v_1$, $v_3$, $v_5$, $v_9$ may execute excessive control to improve their own objectives, thus a coordination among them is needed to alleviate the ECE; meanwhile other agents' impact on reaching the consensus is minor.

\subsection{Numerical experiment}

\begin{figure*}[thpb]
    \centering
    \subfigure[Convergence time]
    {
        \includegraphics[width=0.35\linewidth]{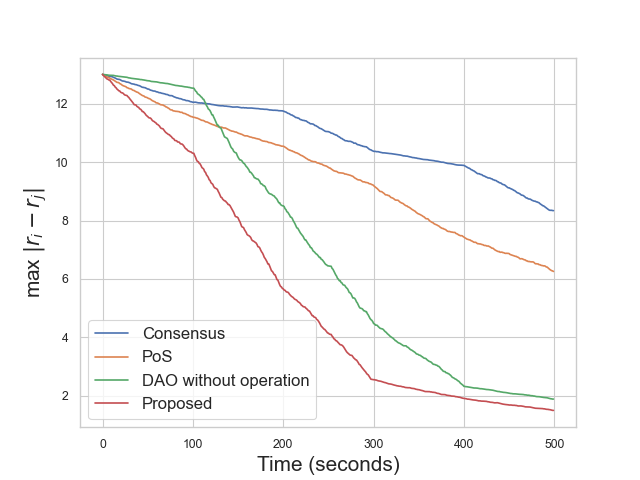}
        \label{converge}
    } 
        \subfigure[Evaluation criteria value]
    {
        \includegraphics[width=0.35\linewidth]{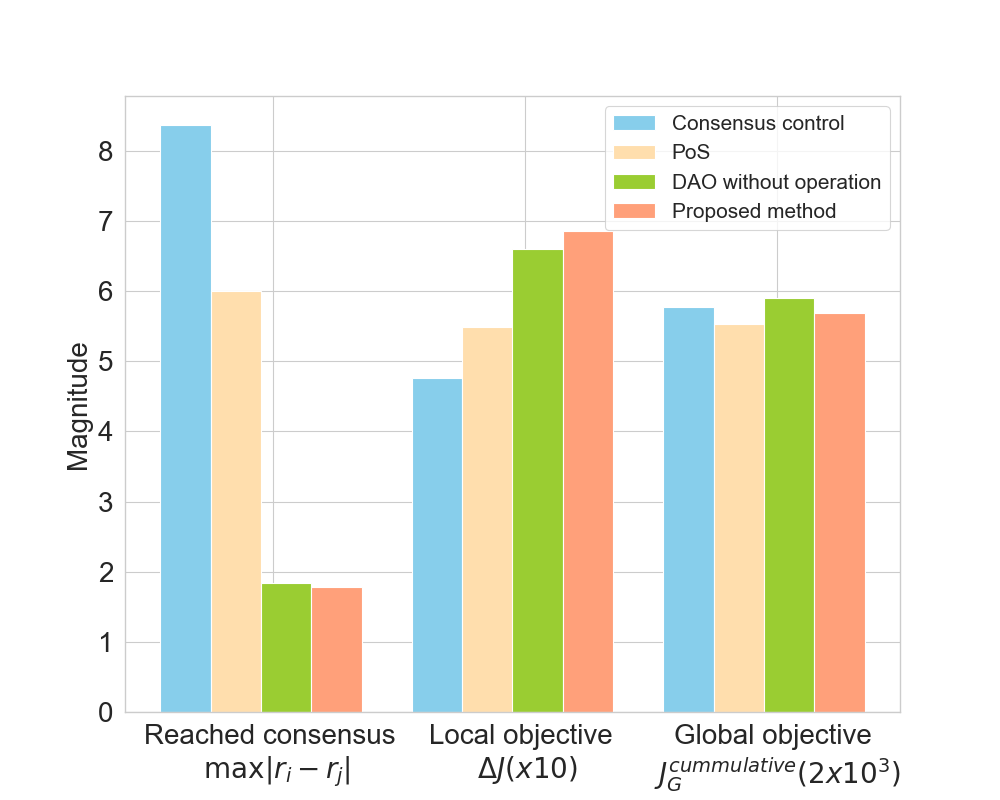}
        \label{value}
    }
    \caption{Experiment results. (a) presents the convergence curve of the maximum discrepancy in ECE among all agents, which are calculated by $\max|r_i-r_j|, \forall i,j \subseteq N$. (b) presents results of three evaluation criteria: (1)The reached consensus on ECE at the end of simulation. (2) The accumulated magnitude of improved local objectives, which are represented by $\Delta J$, that $\Delta J = \sum_{i}(g_{i}(t_{end})-\sum_{i}g_{i}(t_{start}))$. (3) The cumulative ECE, which is calculated by Eq. (\ref{cummulative obj}). Note that infinity values are excluded.}
    \label{result}
\end{figure*}
A numerical experiment is designed in this section to present the feasibility and efficiency of the proposed consensus and incentive mechanism and the operation of DAO. In this experiment, the traditional decentralized consensus control, the PoS-based consensus control, the DAO based control without operation on DAO, as well as the proposed method are examined.  

Similar to Fig. \ref{scenario}, a total of 10 intelligent agents in four different scales are simulated, that $V=\{v_1,...v_{10}\}$, the adjacent matrix of the simulated system is:

\begin{equation}
\small{
\setlength{\arraycolsep}{1.3pt}
    A=
    \begin{bmatrix}
    0 & 0 & 0.1 & 0 & 0 & 0.2 & 0 & 0 & 0 & 0\\
    0 & 0 & 0.3 & 0.1 & 0.1 & 0 & 0 & 0 & 0 & 0\\
    -0.2 & 0.2 & 0 & 0 & 0.1 & 0 & 0 & 0 & 0.03 & 0.1\\
    0 & -0.1 & 0 & 0 & 0 & 0.5 & 0.2 & 0 & 0 & 0\\
    0 & 0 & -0.1 & -0.03& 0 & 0 & 0 & 0 & 0.4 & 0\\
    -0.02 & 0 & 0 & 0 & 0 & 0 & 0 & 0 & 0 & 0\\
    0 & 0 & 0 & 0.2 & 0 & 0 & 0 & 0.4 & 0 & 0\\
    0 & 0 & 0 & 0.1 & 0.2 & 0 & -0.1 & 0 & 0.3 & 0\\
    0 & 0 & -0.1 & 0 & 0 & 0 & 0 & 0.2 & 0 & 0\\
    0 & 0 & 0 & 0.05 & 0 & 0 & 0 & 0 & 0 & 0\\
    \end{bmatrix}
}
\end{equation} 
For simplicity, the detailed system dynamic of each agent is not modeled, while hypothetical functions are adopted to represent their system dynamics, as in Eq. \ref{system dynamic}.
\begin{equation}
\label{system dynamic}
f(x_i, \textbf{u}_i)= x_i \cdot \sin{i}  + \textbf{u}_i \cdot \cos{i}    
\end{equation}
with $u_i \subseteq \{-3, 3\}$. The local objectives $g_i\{x_i,u_i\}$ are designed as:
\begin{equation}
    g_i\{x_i,u_i\}=i \cdot \sin{(x_i)} + u_i^2 \cdot \cos{i} 
\end{equation}
For the response time of each agents and the operation on DAO, $\tau_1$ is 10 seconds, $\tau_2$, $\tau_3$ are 2 seconds, $\tau_4$ to $\tau_6$ are 1 seconds, $\tau_7$ to $\tau_{10}$ are 0.5 seconds, and $\tau_o$ is 5 seconds. In addition, for the decentralized consensus control, $\alpha_i$ and $\beta_i$ are time-invariant coefficients that $\alpha =2$ and $\beta = 1$. For the PoS-based control, $\gamma_i$ is time-invariant that $\gamma_i=1$. In addition, $k_1 = 2$, $k_2 = 5$, $k_3 = 0.3$, $k_4 = 0.1$, and $\Psi = 2$, $\Phi = 4$. A total time of 500 seconds are simulated. The initial state $x_i$ are randomly selected in $[-20,20]$ and the initial control effort $u_i$ are randomly selected in $[-3,3]$. 

The performances of four considered methods are presented in Fig. \ref{result}. It can be observed from Fig.\ref{converge} that the proposed consensus and incentive mechanism, together with the operation on DAO boosts the speed of achieving consensus significantly. In addition, by operating the DAO reiteratively, the convergence speed is able to improve around 20\% comparing to the DAO-based control without such an operation. 

It is observed from Fig. \ref{value} that the reached consensus on the ECE is significantly improved by applying the proposed integrated control method. Additionally, the proposed method achieved a significant improvement in the accumulated value of local objectives by around 50\% comparing to the PoS-based control. Although a compromise of the overall ECE can be observed when applying the proposed method, the compromise is negligible. 

The evaluation results suggest that the cooperation and synchronization of heterogeneous agents in ITS can be extensively achieved by the proposed consensus and incentive mechanism. Specifically, the local objectives can be significantly improved without compromising global objectives, which is realized through balancing the ECE. The performance can be further improved by operating the DAO, which suggests that the capability of DAO-based control in ITS can be enhanced via the proposed operation method. Consequently, the frequency of altering code in the deployed smart contract or deploying new smart contracts can be ultimately reduced with a growing complexity of ITS.           
\section{CONCLUSION}
\label{conclusion}

This paper presents a DAO-based control method that can realize an integrated control of plentiful heterogeneous agents in a complex system, which is applicable in ITS. Specifically, a consensus and incentive control protocol and an operation algorithm on DAO are proposed. By extending the PoS based consensus control protocol, the proposed control protocol aims to increase the cumulative energy consumption efficiency of all intelligent agents within the system. With the consideration of incentives from the rewards of approaching a consensus and the 'support/objection' from connected agents, the system is able to achieve a consensus with improved local objectives. In addition, a heuristic algorithm is proposed to identify critical agents within DAO, in order to extend the capability of the proposed control protocol.         

The experiment results show that the controlled system can achieve a faster consensus, resulting in a higher improved value of local objectives with minor compromise in the global objective, compared to existing decentralized control methods. These results indicate that the proposed method can be applied in practice, with an extensive significance in improving the control performance and extending the control capability of DAO. Moreover, the proposed DAO operation algorithm can be adopted as a quantitative measurement of the DAO-based control's capability, which shows its potential as a guidance in developing the DAO-based integrated control system with ITS becoming growing complex. 

However, the proposed method is examined by a numerical experiment, whereas its performance in practice needs to be evaluated in a traffic simulation platform, with the consideration of several practical issues, such as communication failures. Further, the proposed method is based on the assumption of static connection intensities known a-priori, which can be time-variant in practice. These issues can be further investigated by future research includes applying the proposed method in a traffic simulation platform, as well as developing a distributed learning algorithm to obtain the time-variant adjacent matrix in ITS. 


%


\bibliographystyle{./bibtex/bib/IEEEtran}
\bibliography{./bibtex/bib/IEEEabrv, ./bibtex/bib/IEEEexample}
%

\end{document}